\documentclass[12pt,a4]{article}
\usepackage{amsmath}
\usepackage{bm}
\usepackage{amsfonts}
\usepackage{amssymb}
\usepackage{graphics}
\usepackage[normal]{caption2}
\usepackage{subfigure}
\usepackage{rotating}
\usepackage{citesort}
\def\be{\begin{equation}}
\def\ee{\end{equation}}
\def\ba{\begin{array}}
\def\ea{\end{array}}
\def\bea{\begin{eqnarray}}
\def\eea{\end{eqnarray}}

\begin{document}
\baselineskip 20pt \setlength\tabcolsep{2.5mm}
\renewcommand\arraystretch{1.5}
\setlength{\abovecaptionskip}{0.1cm}
\setlength{\belowcaptionskip}{0.5cm}
%%%%%%%%%%%%%%%%%%%%%%%%%%%%%%%%%%%%%%%
\pagestyle{empty}
\newpage
\pagestyle{plain} \setcounter{page}{1} \setcounter{lofdepth}{2}
\begin{center} {\large\bf N/Z dependence of balance energy throughout the colliding geometries}\\
\vspace*{0.4cm}

{\bf Sakshi Gautam} and {\bf Rajeev K. Puri}\footnote{Email:~rkpuri@pu.ac.in}\\
{\it  Department of Physics, Panjab University, Chandigarh -160
014, India.\\}
\end{center}

\section*{Introduction}
Heavy-ion reactions are unique means to produce in terrestrial
laboratories hot neutron-rich matter similar to those existing in
many astrophysical situations. The rapid advances in technologies
to accelerate radioactive beams provide a great opportunity to
explore the properties and equation of state (EOS) of such
asymmetric neutron-rich matter. As a result of such advancements,
studies on the role of isospin degree of freedom have recently
attracted much attention. These detailed studies of isospin degree
of freedom in nuclear reactions provide valuable probes of the
different formulation of nuclear EOS and specially its
isospin-dependent part. Therefore, the new challenge is to
determine the EOS of asymmetric nuclear matter and in particular
the symmetry energy. In recent past a large number of variables
have been proposed which show sensitivity to symmetry energy
\cite{sym1}. For example, in the low density region, i.e. at
density below the saturation density, isospin diffusion,
isoscaling parameter, neutron to proton ratio, neutron skin
thickness have been found to be sensitive to symmetry energy. On
the other extreme, in high density region, at densities above the
saturation density, $\pi^{+}$/$\pi^{-}$ ratio, neutron-proton
differential flow show sensitivity to symmetry energy. In our
recent study \cite{gaum1}, we show the sensitivity of collective
transverse in-plane flow to the symmetry energy in the Fermi
energy region and its insensitivity in high energy region.
Motivated by this result, in the present work, we study the N/Z
dependence of balance energy for various systems throughout the
colliding geometry range, i.e from central to peripheral one. The
study is carried out within the framework of isospin-dependent
quantum molecular dynamics (IQMD) model. For the details of the
model, the reader is referred to Ref \cite{hart98}.

\section*{Results and discussion}
We simulate the reactions of Ca+Ca, Ni+Ni, Zr+Zr, Sn+Sn, and Xe+Xe
with N/Z varying from 1.0 to 2.0 in small steps of 0.2. In
particular we simulate the reactions of $^{40}$Ca+$^{40}$Ca,
$^{44}$Ca+$^{44}$Ca, $^{48}$Ca+$^{48}$Ca, $^{52}$Ca+$^{52}$Ca,
$^{56}$Ca+$^{56}$Ca, and $^{60}$Ca+$^{60}$Ca; $^{56}$Ni+$^{56}$Ni,
$^{62}$Ni+$^{62}$Ni, $^{68}$Ni+$^{68}$Ni, $^{72}$Ni+$^{72}$Ni, and
$^{78}$Ni+$^{78}$Ni; $^{81}$Zr+$^{81}$Zr, $^{88}$Zr+$^{88}$Zr,
$^{96}$Zr+$^{96}$Zr, $^{104}$Zr+$^{104}$Zr, and
$^{110}$Zr+$^{110}$Zr; $^{100}$Sn+$^{100}$Sn,
$^{112}$Sn+$^{112}$Sn, $^{120}$Sn+$^{120}$Sn,
$^{129}$Sn+$^{129}$Sn, and $^{140}$Sn+$^{140}$Sn; and
$^{110}$Xe+$^{110}$Xe, $^{120}$Xe+$^{120}$Xe,
$^{129}$Xe+$^{129}$Xe, $^{140}$Xe+$^{140}$Xe, and
$^{151}$Xe+$^{151}$Xe at b/b$_{max}$ = 0.2 - 0.4, 0.4 - 0.6 and
0.6 - 0.8. We also use a soft equation of state along with the
standard isospin- and energy-dependent cross section reduced by
  20$\%$, i.e. $\sigma$ = 0.8 $\sigma_{nn}^{free}$.

\begin{figure}[!t]
\centering
 \vskip -1.cm
\includegraphics[angle=0,width=7cm]{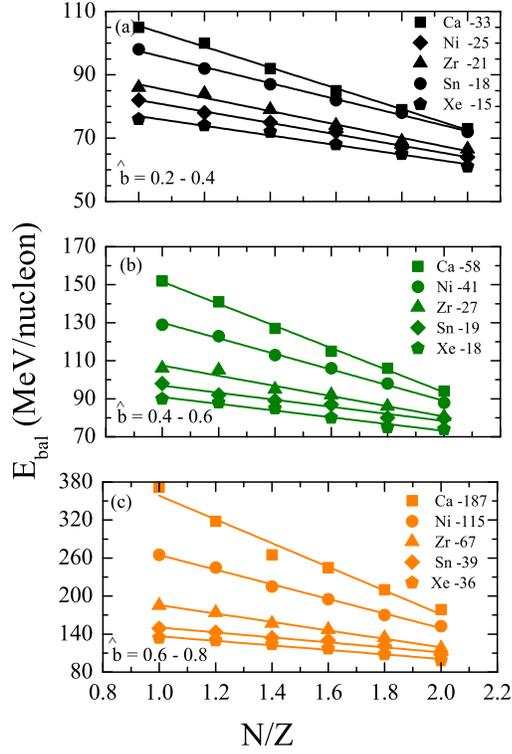}
 \vskip -0.8cm \caption{ N/Z dependence of E$_{bal}$ for various systems. Lines are linear fit. Top, middle and bottom panels represent the results for b/b$_{max}$ = 0.2 - 0.4, 0.4 - 0.6 and
0.6 - 0.8, respectively. }\label{fig1}
\end{figure}

The reactions are followed till the transverse in-plane saturates.
In the present study we use the quantity "\textit{directed
transverse momentum $\langle p_{x}^{dir}\rangle$}" which is
defined as
\begin {equation}
\langle{p_{x}^{dir}}\rangle = \frac{1} {A}\sum_{i=1}^{A}{sign\{
{y(i)}\} p_{x}(i)},
\end {equation}
where $y(i)$ and $p_{x}$(i) are, respectively, the rapidity and
the momentum of the $i^{th}$ particle. The rapidity is defined as
\begin {equation}
Y(i)= \frac{1}{2}\ln\frac{{\vec{E}}(i)+{\vec{p}}_{z}(i)}
{{\vec{E}}(i)-{\vec{p}}_{z}(i)},
\end {equation}
where $\vec{E}(i)$ and $\vec{p_{z}}(i)$ are, respectively, the
energy and longitudinal momentum of the $i^{th}$ particle.  In
figure 1 we display the N/Z dependence of E$_{bal}$ for
b/b$_{max}$ = 0.2-0.4 (top panel), 0.4-0.6 (middle panel) and 0.6
-0.8 (bottom panel). From figure, we find that at all the
colliding geometries E$_{bal}$ follows a linear behaviour with
N/Z. The slopes are 33, 25, 21, 18, and 15 (at b/b$_{max}$ =
0.2-0.4), 58, 41, 27, 19, and 18 (at b/b$_{max}$ = 0.4-0.6) and
187, 115, 67, 39, and 36 (at b/b$_{max}$ = 0.6-0.8)
for the series of Ca, Ni, Zr, Xe and Sn, respectively.  From figure, we find that \\
(i) the N/Z dependence of E$_{bal}$ is steeper for the lighter
systems as compared to the heavier systems at all the colliding geometries, \\
(ii) for a particular isotopic series, the N/Z dependence of
E$_{bal}$ is more at peripheral colliding geometry.\\
(iii) and the change in slope is more for lighter systems as
compared to the heavier systems when we move from central to
peripheral colliding geometries. From figure, we see that for Ca
series, slope increases by almost 400\% when we move from central
to peripheral collisions, whereas for Xe series increase in slope
is almost 150\%.

\section*{Acknowledgments}
 This work has been supported by a grant from Centre of Scientific
and Industrial Research (CSIR), Govt. of India.


\begin{thebibliography}{999}
\bibitem{sym1} M. B. Tsang \emph{et al}., Phys. Rev. Lett. \textbf{102}, 122701
(2009);  B. A. Li, Phys. Rev. Lett. \textbf{88}, 192701 (2002).

\bibitem{gaum1} S. Gautam \emph{et al}., Phys. Rev. C
\textbf{83}, 034606 (2011).

\bibitem{hart98} C. Hartnack \emph{et al}., Eur. Phys. J. A
\textbf{1}, 151 (1998); R. K. Puri \emph{et al}., Nucl. Phys. A
\textbf{575}, 733 (1994); E. Lehmann \emph{et al}., Phys. Rev. C
\textbf{51}, 2113 (1995); R. K. Puri \emph{et al}., J. Comp. Phys.
\textbf{162}, 245 (2000).
\end{thebibliography}
\end{document}